\documentclass[10pt,journal,compsoc]{IEEEtran}
\usepackage[utf8]{inputenc}
\usepackage{tabularx}
\usepackage{graphicx,color}
\usepackage{balance}
\usepackage{array}
\usepackage{varwidth}
\usepackage{pgfplots}
\usepackage{comment}
\usepackage[normalem]{ulem}

\newcolumntype{P}[1]{>{\centering\arraybackslash}p{#1}}

\title{Design Guidelines for Blockchain-Assisted 5G-UAV Networks}

\author{
\IEEEauthorblockN{\textbf{Moayad Aloqaily},
\textit{Member, IEEE}, \textbf{Ouns Bouachir}, \textit{Member, IEEE}, \textbf{Azzedine Boukerche}, \textit{Fellow, IEEE}, \textbf{Ismaeel Al Ridhawi}, \textit{Senior Member, IEEE}}

\IEEEcompsocitemizethanks{

\IEEEcompsocthanksitem M. Aloqaily is with Al Ain University, UAE. \protect E-mail: maloqaily@ieee.org

\IEEEcompsocthanksitem O. Bouachir is with College of Technological 
Innovation, Zayed University, UAE. \protect E-mail: ouns.bouachir@zu.ac.ae

\IEEEcompsocthanksitem A. Boukerche is with University of Ottawa, Ottawa, ON, Canada. \protect E-mail: boukerch@uottawa.ca

\IEEEcompsocthanksitem I. Al Ridhawi is with Kuwait College of Science and Technology, Kuwait. \protect E-mail: i.alridhawi@kcst.edu.kw
}
}

\begin{document}

\maketitle

\begin{abstract}

Fifth Generation (5G) wireless networks are designed to meet various end-user Quality of Service (QoS) requirements through high data rates (typically of Gbps order) and low latencies. 
Coupled with Fog and Mobile Edge Computing (MEC), 5G can achieve high data rates, enabling complex autonomous smart city services such as the large deployment of self-driving vehicles and large-scale Artificial Intelligence (AI)-enabled industrial manufacturing. 
However, to meet the exponentially growing number of connected IoT devices and irregular data and service requests in both low and highly dense locations, the process of enacting traditional cells supported through fixed and costly base stations requires rethought to enable on-demand mobile access points in the form of Unmanned Aerial Vehicles (UAV) for diversified smart city scenarios.
This article envisions a 5G network environment that is supported by blockchain-enabled UAVs to meet dynamic user demands with network access supply. The solution enables decentralized service delivery (Drones as a Service) and routing to and from end-users in a reliable and secure manner. Both public and private blockchains are deployed within the UAVs, supported by fog and cloud computing devices and datacenters to provide wide range of complex authenticated service and data availability. Particular attention is paid to comparing data delivery success rates and message exchange in the proposed solution against traditional UAV-supported cellular networks. Challenges and future research are also discussed with highlights on emerging technologies such as Federated Learning.
\end{abstract}


\section{Introduction}
The rapid growth in communication traffic between Internet of Things (IoT) devices has revolutionized the current communication framework, leading to smaller cell architectures and ultra-dense networks (UDNs). The fifth generation (5G) network was designed to meet stringent user quality of service (QoS) requirements in terms of higher data rate and low latency. Many smart city applications have been deployed as a result of this ever-evolving communication technology. For instance, smart intelligent autonomous self-driving and service provisioning vehicles are now capable of achieving the intended tasks with the aid of fog and cloud computing \cite{ref1}. Most of the time-sensitive critical decisions can now be made at the fog or the resource-rich edge without full reliance on the cloud. Such an approach introduced the concept of fog-to-cloud (F2C) and fog-to-fog (F2F) computing \cite{N5}.

\begin{figure*}[!h]
    \centering
    \includegraphics[scale=0.4]{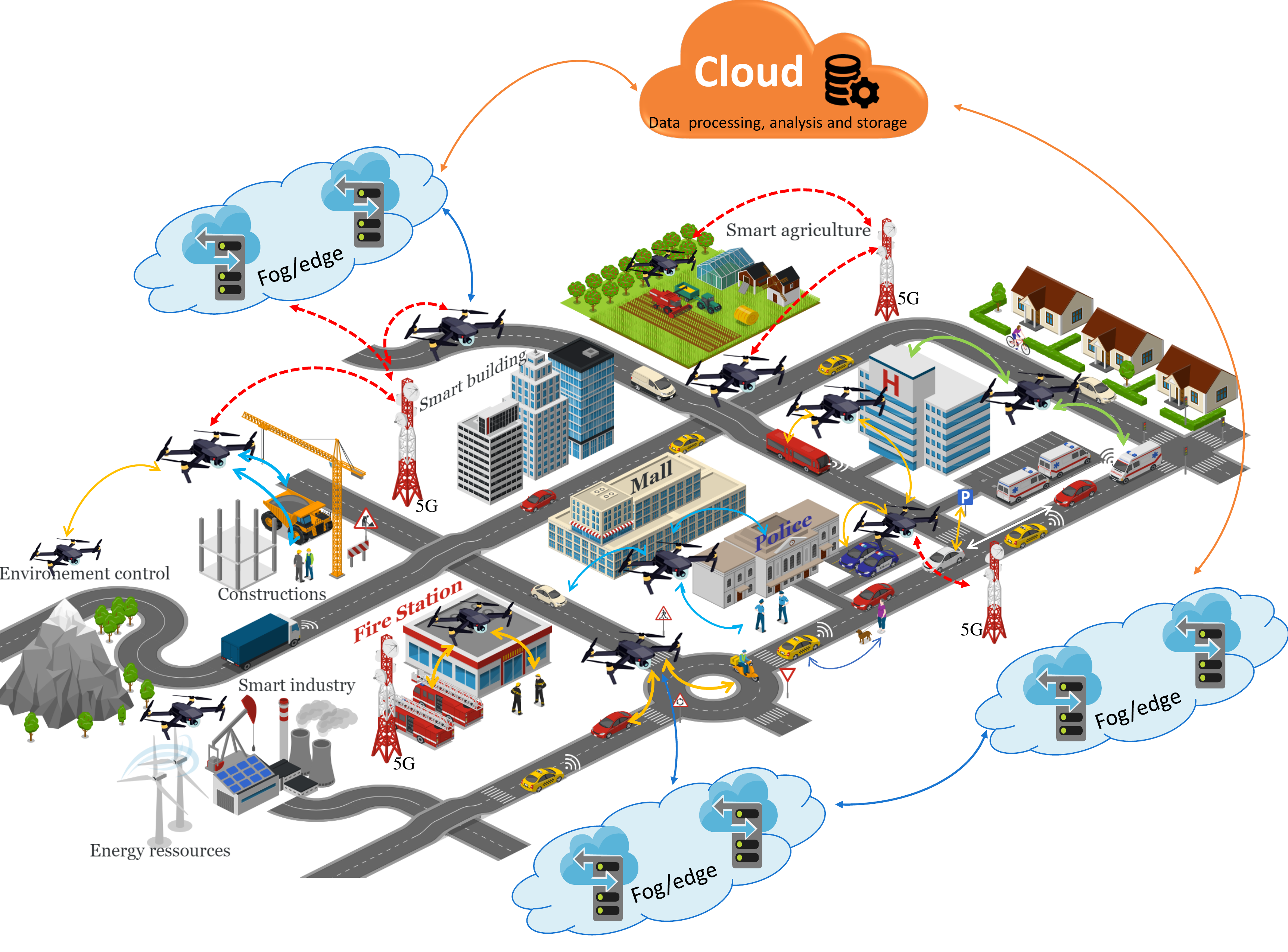}
    \caption{An Overview of the Blockchain-Assisted 5g-Drone Network Scenario}
    \label{fig:general}
\end{figure*}

However, in today’s smart city environments, and with the ever-growing numbers in IoT devices, especially in UDN environments, 5G networks will face limitations due to unpredictable and non-uniform loading \cite{ref3}. As such, if the supplied network capacity does not match the throughput demand, end-user QoS requirements will not be met. To solve such an issue in regards to the disproportion between the irregular demand and the limitations in access availability, UDN could be assisted through mobile serving stations, namely, unmanned aerial vehicles (UAVs), or simply drones. UAV mobile base stations (BSs) will improve the QoS of the wireless network by increasing network capacity and improving the coverage area \cite{ref4}\cite{DBLP:journals/cem/GargAKB19}. Such mobile BSs can be moved to highly dense locations close to the end-users. Drones also have the potential of replacing or complementing, to say the least, cellular networks in high interference conditions or locations where it is not economically feasible to install a permanent infrastructure. For instance, in cases where there are few active end-users, deploying UAVs may sufficiently serve the users’ needs without reliance on high-end cellular BSs. Moreover, running a few drones in those locations might reduce the overall power consumption (i.e. by switching on and off the drones according to the number of active requests compared to an all-time active fixed cellular BS).

Although mitigating the disproportion between irregular end-user requests and the available 5G access through an increase in the number of fixed and mobile stations, such over-provisioning leads to more complex service provider management issues. To this end, we investigate the feasibility and significance of a novel blockchain-assisted 5G-UAV network over other state-of-the-art UAV-supported 5G networks. The solution relies on a decentralized service delivery and routing technique to and from end-users through a 5G-supportive UAV network. Both public and private blockchains are deployed in a smart city to ensure secure communication between participants. Mobile UAVs are deployed in ultra-dense areas suffering from network congestion and decreased QoS to provide efficient and reliable data routing. UAVs communicate with both fog/edge devices and cloud datacenters to ensure extensive coverage of a multitude of smart city services and access to resources. In addition, we identify challenges associated with the blockchain-assisted 5G-UAV network such as energy consumption and reliability. We demonstrate through simulation results that the proposed solution can be beneficial, in terms of data delivery success and minimal data exchange.

\section{Drones as a Service (DaaS)}
\subsection{Operations}
With the remarkable fast advances in AI and machine learning (ML), extraordinary capabilities are designed empowering the concept of smart autonomous systems such as robots and autonomous vehicles.
UAVs, or simply, drones, are another example of systems that may benefit from these advances. Used for the first time in the second world war by the American army, in an attempt to use a manned aircraft in an unmanned manner, the name ``drones'' have always been coupled with  military. Nowadays, UAVs are available with diverse sizes and capabilities and used in civil domains for various’ applications. They perform at different automation levels, ranging from fully controlled (no automation), to fully autonomous that can be reflected by a swarm of drones co-operating together to fulfill a common mission.

As shown in Figure \ref{fig:general}, UAVs can be  active actors in all types of smart city services and applications. They can be used with police and fire stations for emergency situations, as well as in smart transport management,  healthcare, agriculture, industry, smart grid, environment control, and much more.

Coupled with the most advanced paradigms of communication (i.e. 5G), data storage (using edge, fog, cloud, blockchain technologies) and processing (e.g. ML and AI), DaaS is able to provide a big range of smart services (figure \ref{fig:general}). Indeed, drones are able to perform any task, on-demand, using the different communication infrastructures deployed around the city. The various types of collected data, the users’ requests, or the transactions among the service providers can be processed and stored by edge or cloud computing, allowing the extraction of valuable information that can be very useful in enhancing quality of experience, safety  and life in the smart city.

\subsubsection{Services}

Thanks to their 3D movement flexibility, UAVs do not have complex traversing restrictions as seen in roads. They can move in various directions and fly over inaccessible and dangerous locations. Various applications can benefit from this feature. Those applications can be grouped into three different classes based on the task allocated to the aircraft:
\begin{itemize}
  
\item \textbf{Flying IoT device.}
Equipped with diverse types of sensors (e.g. gas and temperature sensors), a drone can reach inaccessible and dull locations and get better visibility than any other IoT device to collect accurate data about a situation. For instance, drones can fly around big buildings, mountains, bridges, hazardous areas (e.g. after natural disasters), etc. An Example of sensed information the weather forecast information, which can be forwarded, in real-time, to the closest edge device, to the cloud, or stored in the drone's memory and collected by the ground station at the end of the exploratory trip. UAVs are considered a cost-effective flying IoT devices used to collect truthful information for various applications.  Surveillance, monitoring, environment measurements are popular examples of IoT applications-based drones. 
\item \textbf{Mobile flying base station}
Due to the expansion in the number of connected devices, especially in crowded areas, communication networks are facing many challenges to meet all these simultaneous requests. Many researchers have focused on this issue and have proposed to add mobile small cells (MSC) that can support the connection provided by the mega-cell. \cite{PIMRC}
Flying few meters above the users, UAVs offer an important coverage area that can be provided by a flying mobile small cell. Also, UAVs can be used as base stations or relays to connect devices in areas where no communication infrastructure is used (p2p networks, rural areas and after hazardous events).

\item\textbf{Flying robots}
Advances in technology are creating intelligent systems based on unprecedented paradigms. AI and ML are creating a revolution in all technological sectors by providing cutting-edge devices such as robots that are able to collect information about their surrounding environment, analyze the current situation, take decisions and perform various tasks in real-time. UAVs are considered flying robots that, based on innovative algorithms, can replace humans in many dull missions \cite{Tony_NYU}. Examples are the amazon delivery and flying ambulance projects.
\end{itemize}
\subsubsection{UAV Automation Levels}
Drones can perform various operations and tasks under different levels of automation that can go from fully controlled drones to fully autonomous. The level of automation is defined by how the drone can monitor the progress of its mission.

\begin{itemize}
    \item\textbf{Zero automation level}: A pilot can control the drone’s operations from the ground by sending instructions for each movement and action. In this situation, the aircraft should obey all the received directions passively.  
    \item \textbf{Hybrid automation level}: the UAV’s automation level can increase progressively, reducing the human interposition during the mission. Some decisions may be taken locally by the drones, based on surrounding events, while some others are received from the ground controller.
    \item \textbf{Full automation level}: based on advances in AI and ML algorithms, drones have full control of their operations including trajectory planning, sudden movement and data exchanging with the ground station. To make fast and wise decisions, drones should rely on their sensors and the data received from the surrounding (e.g. other drones, vehicles, infrastructure, etc.).
\end{itemize}

Drones having a certain automation level, and may perform as a swarm of cooperative agents \cite{bouachir2019testbed}\cite{8648452}. By continuously exchanging data between the drones, they can create an idea about the situation and update their tasks based on that. For instance, each drone may change its path plan based on other aircraft trajectories or task results.
\subsection{UAV Networks Challenges}
As explained above, all UAV operations are based on a continuous exchange of data between the drone and various surrounding devices such as other drones, base stations, vehicles, infrastructure, etc. 
A drone exchanges diverse types of messages:  information related to its task (e.g. geographic location or pilot instructions), gathered IoT data, and much more \cite{liang2017towards}. These messages can be generated by the drone, sent to the drones or simply relayed to other devices through the drone. The communication system used by the drones faces several challenges:

\begin{itemize}
    \item \textbf{Fast delivery}: To guarantee the best task performance, drones should exchange information about surrounding events (or pilot instructions) on time, with the shortest delay, to make instant reactions. Also, some of the forwarded IoT data may be critical and require low transmission delay.
    \item \textbf{Trust and privacy}: drones may perform as a swarm or individually and they may relay information exchanged between other devices. The privacy of the exchanged messages, the participating aircraft identity verification and authentication are critical issues that should be taken into consideration.
    \item \textbf{Security}: Drones, operating in the sky, communicate thorough wireless communication technologies, and may be vulnerable to several privacy and security risks. Drones can be easily controlled by hackers to manipulate their tasks, to send wrong data and to target data accountability, data integrity, data authorization, and reliability.
    \item \textbf{QoS}: drones exchange various types of data such as video surveillance (real-time traffic) and emergency events that require immediate reactions \cite{bouachir2019testbed}. These messages have different requirements in term of QoS such as throughput, reliability and delay, that may change with time.
    \item \textbf{Energy consumption}: Most of the drones are equipped with limited batteries that have short lifetime. The communication exchanges, the computing processes running on the UAV and the carried payload increase the energy consumption and reduce the lifetime of the UAV. 
\end{itemize}

\begin{table*}[!h]
    \centering
    \label{table1} 
    \caption{Comparison between different drone communication platforms. $\uparrow$ \textit{means comparatively high}, $\downarrow$ \textit{means comparatively low}}
    \begin{tabular}{|P{2cm}||P{4.5cm}|P{4.5cm}||P{4.5cm}||}
    \hline
   \textbf{Characteristics}&\textbf{Basic UAV Networks}& \textbf{5G UAV Networks} &  \textbf{BC Assisted 5G UAV Networks} \\
         \hline
         \hline
         \textbf{Communication range}& Limited range that depends on the used technology (i.e. Zigbee or Wifi)&Extended range based on 5G infrastructure&Extended range based on 5G infrastructure\\
         \hline
         \textbf{QoS }&Fair data delivery speed with limited throughput specially with Zigbee that cannot support video messages & Very fast data delivery with high throughput and reliability level&Very fast data delivery with high throughput and reliability level\\
         \hline
         \textbf{Identity verification}&No UAV identity verification&No UAV identity verification& UAV identity is checked so not any drone can participate in the mission\\
         \hline
         \textbf{Data security}& $\downarrow$  & $\downarrow$  & Data is stored in BC that provides security\\
         \hline
         \textbf{Privacy} &$\downarrow$ &$\downarrow$ &Data is stored in BC that provides privacy\\
         \hline
         \textbf{Trust} & $\downarrow$  &$\downarrow$ & Trust is provided since all participant's identities are verified, their privacy is guaranteed and data storage is secure\\
         \hline
         \textbf{Decentralization}&Centralized system&Centralized System&Decentralized System\\
         
         \hline
        \textbf{ Resource management} &Basic, depends on the used technology& 5G resource management& Advanced, managing 5G resources using BC\\
         \hline
         \textbf{Power efficiency}&$\downarrow$&$\downarrow$& Thanks to the resource management features, this system provide UAVs power efficiency \\
         \hline
         \textbf{Scalability} &$\downarrow \downarrow$&$\downarrow$& Better scalability \\
         \hline
         \textbf{Intelligence} &&&\\
         \hline
         \textbf{UAV movement flexibility} &Restricted, within the communication range &$\uparrow$&$\uparrow$\\
         \hline
         \textbf{Autonomy}&Limited&Limited& Very advanced\\
         \hline
         \textbf{Swarm operations} &Limited&Limited&Advanced\\
         \hline
    \end{tabular}
   
\end{table*}

\section{BlockChain for Drones}
\subsection{BlockChain Overview}
Generally speaking, blockchain (BC) is a distributed and tamper resistant ledger that does not rely on a centralized authority to establish trust, with a core layer mechanism for decentralized trust management. Since the introduction of Bitcoin, blockchain is used in applications such as cryptocurrency, secure storage and asset transfer.

The simplest form of blockchain is a linked list of ``blocks'' that contains various types of information. For example, transactions that record Bitcoin money transfers. Each participant stores a local blockchain, and uses some form of a ``consensus'' mechanism to establish its exact order and content. Popular mechanisms include proof-of-work (PoW), proof-of-stake (PoS) and practical Byzantine-fault-tolerance (PBFT), each of which has different trade-off and suitable usage scenarios \cite{tseng2020blockchain}\cite{Aloqaily2020}. Blockchain technology has been envisioned because of its robustness in providing trust and anonymity in any commodity trading such as energy \cite{N4}.

Each block is ``chained'' to the previous one using a cryptographical data structure called a hash pointer, and participants jointly verify all transactions by examining the content and the accompanying hash pointers. Thus, false transactions are rejected if enough participants are correct. Such mechanisms ensure the system is tamper-proof, since an adversary cannot persuade correct participants to switch to an incorrect branch of the blockchain. This is also how blockchain intuitively establishes decentralized trust \cite{bera2020designing}\cite{N6}.

The original intention of blockchain was to create a public network similar to the models of Bitcoin, Ethereum and Litecoin. But due to performance limitations private blockchains have also been developed. The difference between public and private blockchains depends on the permissions of the participants. Anyone is free to join a public blockchain, whereas private versions require an authority to approve membership. Therefore, public types are known as permission-less, and private types as permissioned. The benefit of identity management is that private blockchains are usually an order of magnitude faster and transaction costs are lower due to the reduced number of processing nodes. The drawback of blockchain is that identity/permission management is not as scalable and decentralized. 

There are also hybrid solutions that combine private and public blockchains, referred to as ``consortium blockchains''.

\subsection{BlockChain for Drones and 5G}
\subsubsection{Drones and 5G}
With the deployment of 5G, many services can benefit from the advantages of this technology in terms of data rate and energy consumption. 5G and UAVs have a mutual relationship. Each technology can provide a range of advantages to the other that may have a significant impact on the quality of various smart city services:
\begin{itemize}
    \item \textbf{5G in service of UAVs}: As mentioned previously, the biggest challenges for UAV networks are fast data delivery, QoS and energy consumption. 5G is designed to provide these features: highest data rate, better QoS and it was proved that it is an energy-aware technology. This makes 5G a promising solution to make UAVs communicate together and with the different smart city’s components.
    \item \textbf{UAVs in service of 5G}: cellular networks suffer from their signal quality degradation due to several reasons such as interference, vast number of connected devices, big buildings and their thick metallic infrastructure \cite{ PIMRC}. Also, some areas may be outside of the cellular coverage like rural areas or after hazardous situations. The fast growing IoT traffic is very challenging for static base stations in 5G networks. Thanks to their flexible mobility and low-cost, UAVs can serve as MSC that enhances the cellular network coverage and provide a support for 5G infrastructure empowering new 5G applications.
\end{itemize}

\subsection{Concept}

Nowadays, communication is no longer traditional. It's not only that we have mobile phones communicating, but they also connected to the cloud, edge, and fog, as shown in Figure \ref{fig:general}. Such a heterogeneous environment is sensitive to many issues; trust is the most important among all. On-demand \textit{trust} in ultra-dense wireless networks is not an easy task. It requires high speed communication, identification, and authentication. Drones are capable of communicating with each other through trust, to relay information, regardless of the owner (e.g. service provider). ML and AI are two techniques that can be used to identify intrusive traffic. Moreover, blockchain can be used to certify the true identity of drones.

This section presents an architecture that secures the utilization of drones as on-demand flying nodes for inter-service operability between multiple service providers by employing the features of machine learning and blockchain.
\begin{figure}
    \centering
    \includegraphics[scale=0.48]{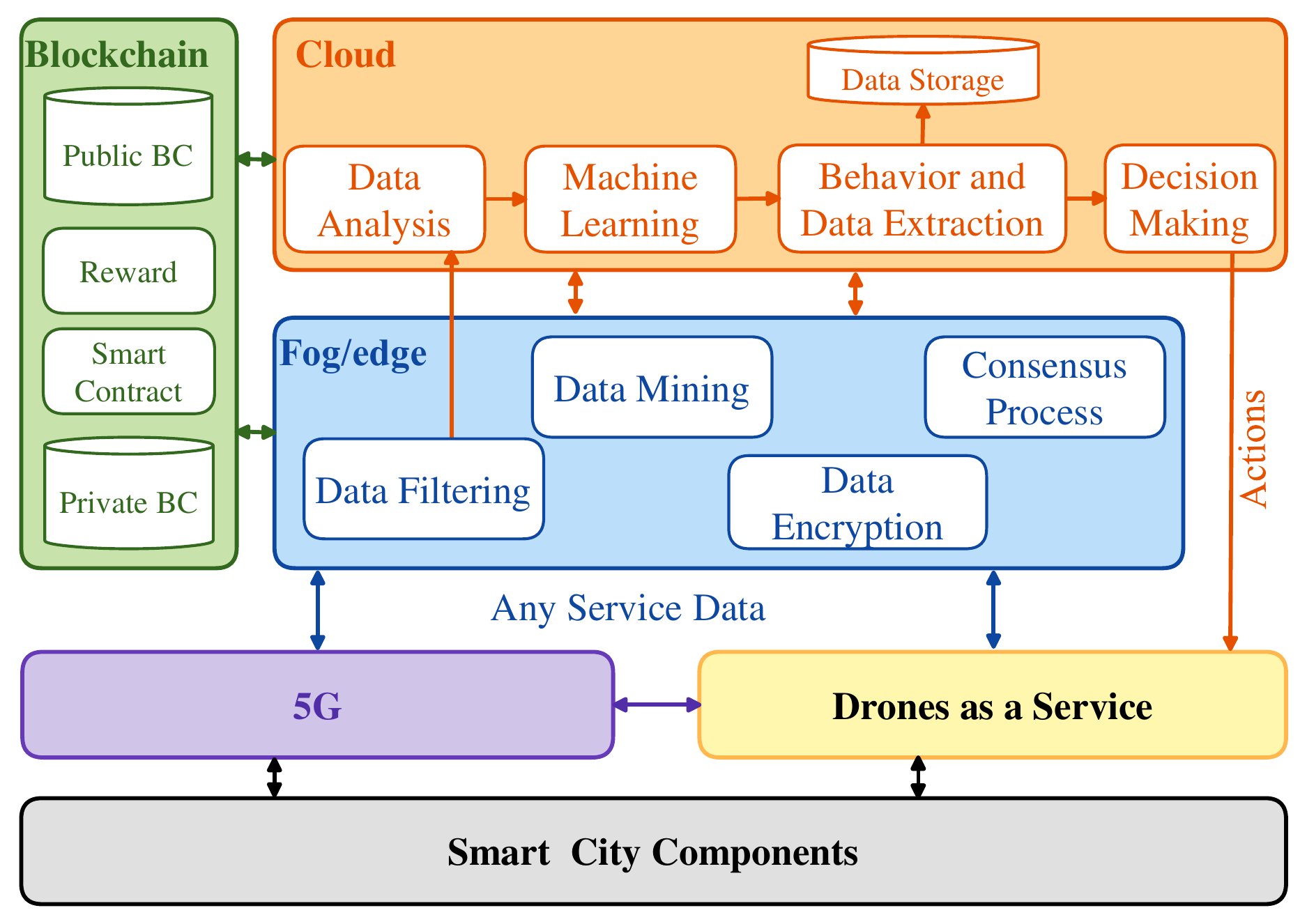}
    \caption{System Architecture.}
    \label{fig:fig2}
\end{figure}

Figure \ref{fig:fig2} presents various components and strategies for utilizing the use of 5G and blockchain-based drones. Further, Table \ref{table1} providers an overview of the characteristics and implications of the blockchain-based drones and compares the deployment issues with the centralized and traditional drone-assisted wireless networks.

Drones are facilitating smart city sub-component services through the 5G infrastructure. Data processing and storage is usually handled at the edge or the cloud. Blockchain can be used to integrate drones with fog and cloud site by insuring data integrity and secure communication. This in return makes the system more reliable and trustworthy.

The use of blockchain-assisted drones in this architecture improves data security across several layers. Blockchain provides a level of security that limits external hacking and data retrieval when a drone physically falls into the wrong hands. Moreover, it validates and ensures the identities of participating UAVs performing a task. Thus, this architecture provides trust among drones and service beneficiaries.

Two types of blockchains can be seen from the architecture and are used according to different scenarios (private or public) where service providers use public or private keys. Public keys are used when the service is visible to all parties while private keys belong to service providers for specific services. Blockchain authenticates communications among UAVs by tracking all their transactions, makes it available to all network nodes, and achieves data integrity (i.e. cryptography) to provide tamper-resistance. The blockchain is decentralized with no centralized authority to access or modify transactions. All network participants must reach a \textit{consensus} to validate transactions in a secure way.

The proposed architecture is secure by the principles of blockchain, where no centralized authentication mechanism such as the public key is used by the UAVs for identifying each other. For example, when providers mutually agree to providing services for a specific area, they also share the pool of secret keys which are the private keys for the drones used to exchange load information while executing a drone's service task. As such, drones and service providers need to execute \textit{drone smart contracts}.  
Blockchain can be used to enhance the performance of the UAV-5G communication network by proposing a spectrum trading platform between the aerial and terrestrial systems allowing a secure spectrum sharing between the network operator’s and UAV service providers. This feature provides privacy based on a distributed sharing information platform and limits the risk of unauthorized spectrum utilization by malicious nodes.

\subsection{Advantages}
The proposed blockchain-assisted 5G-UAV network provides several advantages and improves the QoE of drone services. These benefits are summarized as follows:
\begin{itemize}
    \item Better network QoS in terms of delay, throughput and reliability that are essential requirements in UAV service’s to provide the best QoE.
    \item Data privacy and security based on BC distributed ledger.
    \item Identity of participating drones is verified for each new transaction, providing trust and anonymity that protects from malicious attacks trying to join the UAV network. 
    \item Distributed platform-based immutability, transparency and traceability that increases trustworthiness between agents and enables the deployment of high automation systems.
    \item Better resource and spectrum management that allows to control the interference in UAV-5g-based networks which provide better scalability, accessibility and power efficiency. This advantage allows to create a trusted scheme for interference monitoring in flying small cells.
\end{itemize}

\section{Performance Evaluation: A Scenario}
To test the effectiveness of the proposed model, analysis has been performed through Network Simulator 3 (NS-3). To maintain uniformity, all experiments have been conducted on i7-6500U CPU @ 2.5 GHz with 8 GB of RAM. Simulator settings are summarized in Table II. In the simulation scenario, a test is conducted where up to 100 mobile devices are scattered along a network area of 1500m x 1500m, 20 mobile drones fly over them to form a relay and data delivery platform, and one next generation node B (gNB) base station is placed to cover the entire simulation environment. Communication between UAVs is conducted using a flying ad-hoc network topology (FANET). The UAVs are randomly redistributed to mobile node dense locations throughout the simulation environment at predefined time intervals. Such a technique would represent the movement of UAVs to locations of high-dense mobile nodes to aid in the process of all-time connectivity and faster data delivery. Mobile nodes and UAVs are equipped with two communication interfaces, allowing them to have node-to-drone communication (N2D), node-to-BS communication (N2B), and drone-to-drone communication (D2D). Three solutions were compared, namely, i) traditional node-to-node communication using the BS (N2N-BS), ii) node-to-node communication using both UAVs and BS without blockchain (N2N-UAV without BC), and iii) node-to-node communication using both UAVs and BS with blockchain (N2N-UAV with BC). The tests conducted focused mainly on data delivery success rate and message exchange overhead.

\begin{table}[ht]
\caption{Simulator Settings}\label{table2}
\begin{tabular}{|p{2.75cm}|l|}
\hline
\textbf{Simulation Parameters}       & \textbf{Numerical Values}                                                                                                \\ \hline
\textbf{Communication protocol}      & \begin{tabular}[c]{@{}l@{}}IEEE 802.11n (for N2D and D2D)\\ LTE gNB with 1 Gbps bandwidth\\-(for N2B and D2B)\end{tabular} \\ \hline
\textbf{Area}                        & 1500 m X 1500 m                                                                                                          \\ \hline
\textbf{Number of fixed BSs}         & 1                                                                                                                        \\ \hline
\textbf{Number of UAVs}              & 20                                                                                                                       \\ \hline
\textbf{Number of end-users}         & 10 – 20 mobile nodes                                                                                                     \\ \hline
\textbf{UAVs placement}              & Random – uniform distribution (initially)                                                                                \\ \hline
\textbf{Mobility model}              & Random – waypoint (for mobile nodes)                                                                                     \\ \hline
\textbf{Drone flying heights}        & 50 m                                                                                                                     \\ \hline
\textbf{Simulation Duration} & 60 s                                                                                                                     \\ \hline
\textbf{CBR packet length}           & 512 bytes                                                                                                                \\ \hline
\textbf{CBR packet interval}         & 10 ms                                                                                                                    \\ \hline
\textbf{Transmission power}          & \begin{tabular}[c]{@{}l@{}}0.1 W (for UAVs)\\ 0.01 W (for mobile nodes)\end{tabular}                                     \\ \hline
\end{tabular}
\end{table}

\subsection{Authenticated Data Delivery Success Rate}
\begin{figure}[ht]
\centering
\begin{tikzpicture}
\begin{axis}[
    xlabel={Number of Mobile Nodes},
    ylabel={Packet delivery success rate (\%)},
    legend style={at={(0.02,0.02)},anchor=south west,font=\fontsize{7}{8}\selectfont},
   ymin=50, ymax=100,
    ytick={50,55,60,65,70,75,80,85,90,95,100},
    xtick={10,20,30,40,50,60,70,80,90,100},
    grid style=dashed,
    width = 9cm,
    height = 6.5cm,
    ]
    \addplot[
    color=black!00!teal,
    mark=otimes,
    line width=1pt,
    mark size=2pt
    ]
    coordinates {
    (10,95)(20,92)(30,88)(40,84)(50,80)(60,76)(70,71)(80,66)(90,61)(100,55)
    };
 \addplot[
     color=white!00!red,
     mark=square,
     line width=1pt,
     mark size=2pt,
     ]
     coordinates {
     (10,96)(20,93)(30,89)(40,85)(50,81)(60,77)(70,73)(80,69)(90,65)(100,60)
    }; 
    \addplot[
    color=black!00!purple,
    mark=triangle,
    line width=1pt,
    mark size=2pt,
    ]
    coordinates {
    (10,98)(20,96)(30,95)(40,93)(50,91)(60,89)(70,86)(80,83)(90,80)(100,77)
    };
    \legend{N2N-BS, N2N-UAV without BC, N2N-UAV with BC}
\end{axis}
\end{tikzpicture}
    \caption{Comparing the overall packet delivery success rate among the three different techniques.}    \label{fig:result3}
\end{figure}
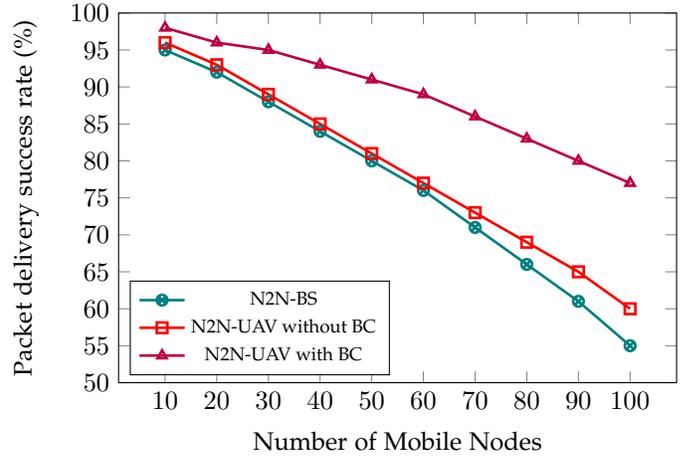

Authenticated data delivery success rate is the percentage of positive responses received for all packets sent from the source mobile node to the destination mobile node. Results depicted in Figure \ref{fig:result3} show that the proposed solution, namely, N2N-UAV with BC outperforms traditional communication techniques and non-authenticated N2N-UAV state-of-the-art solutions. It is important to note that without proper data authentication (with the aid of blockchain), sensitive data may be intercepted by hostile UAVs. From the figure, we see that as the node density increases within the environment, the gap in success rate between the proposed and traditional methods increase. For instance, when comparing N2N-UAV with BC to N2N-BS with 100 mobile nodes in the network, we see that the success rate for N2N-UAV with BC is almost 77\% compared to 55\% for N2N-BS. Moreover, the success rate for node-to-node communication without reliance on blockchain is 60\%. Such results indicate that with high mobility cases, like in UDNs, UAVs are capable of adjusting their location to meet service demands in accordance with mobile node density. For instance, whenever node density increases in a certain location within the network, one or more extra UAVs move to that location to ensure continuous and reliable data delivery. The proposed technique not only delivers data in a timely manner, but also assures that data communication is authenticated to ensure high delivery success rates.

\subsection{Message Overhead}
\begin{figure}[ht]
\centering
\begin{tikzpicture}
\begin{axis}[
    xlabel={Number of Mobile Nodes},
    ylabel={Number of Messages},
    legend style={at={(0.02,0.99)},anchor=north west,font=\fontsize{7}{8}\selectfont},
   ymin=0, ymax=3000,
    ytick={500,1000,1500,2000,2500,3000},
    xtick={10,20,30,40,50,60,70,80,90,100},
    grid style=dashed,
    width = 8.75cm,
    height = 6.5cm,
    ]
    \addplot[
    color=black!00!teal,
    mark=otimes,
    line width=1pt,
    mark size=2pt
    ]
    coordinates {
    (10,500)(20,530)(30,590)(40,650)(50,700)(60,770)(70,870)(80,990)(90,1100)(100,1200)
    };
 \addplot[
     color=white!00!red,
     mark=square,
     line width=1pt,
     mark size=2pt,
     ]
     coordinates {
     (10,1000)(20,1050)(30,1100)(40,1180)(50,1300)(60,1430)(70,1550)(80,1700)(90,1900)(100,2150)
    }; 
    \addplot[
    color=black!00!purple,
    mark=triangle,
    line width=1pt,
    mark size=2pt,
    ]
    coordinates {
    (10,1200)(20,1250)(30,1300)(40,1400)(50,1550)(60,1720)(70,1900)(80,2150)(90,2400)(100,2650)
    };
    \legend{N2N-BS, N2N-UAV without BC, N2N-UAV with BC}
\end{axis}
\end{tikzpicture}
    \caption{Total number of packets exchanged in the network for data delivery between source and destination nodes using the three different data delivery techniques}    \label{fig:result4}
\end{figure}
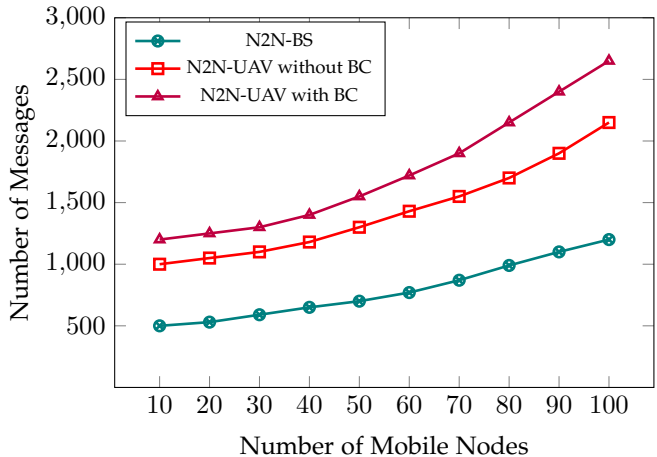

It is well-known that with any short-range communication technique, the number of packets sent from the source to the destination increase in accordance to the number of relay nodes introduced. Moreover, by using blockchain the number of packets increase more to ensure that communication between UAVs is authenticated. Thus, there is always a trade-off between an increase in the communication overhead and system efficiency in terms of reduced delay, power usage reduction and much more. For example, an increase in the network traffic as a result of taking a different data delivery route might speed up the data delivery process if the BS or another primary UAV are busy with handling other data delivery requests. We see from Figure \ref{fig:result4} that when comparing a UAV communication method (N2N-UAV) to a traditional cellular communication method (N2N-BS), the N2N-BS technique outperforms N2N-UAV in terms of reduced network traffic overhead, where the overall number of sent messages is 1200 with 100 mobile nodes network density. When relying on a UAV solution, the overall number of exchanged messages is 2150 for N2N-UAV without BC and 2650 for N2N-UAV with BC. Although results show that a traditional N2N-UAV technique outperforms the our proposed solution that relies on blockchain, the difference in message exchange is modest compared to the gains achieved in terms of authentication and privacy.

\section{Challenges and Future Research}
BC- assisted 5G UAV network empowers drone services by providing security, privacy, trust and better resource management. However, the deployment of this system may face a few challenges:
\begin{itemize}
    \item 	\textbf{Scalability and QoS}: Blockchain faces performance challenges in terms of scalability, throughput and delay due to the rapidly growing volume of data exchanged by such systems. Indeed, data mining requires several verification's performed by each blockchain participant. When the data generation rate is faster than the mining process, the performance of the overall system is affected in terms of throughput and delay.  Scalability is one of the critical challenges causing QoS degradation.
    
    \item 	\textbf{Federated Learning UAVs}: 5G provides a unique communication infrastructure to run sophisticated smart applications of autonomous systems. Flying close to the end-users, drones can be utilized as relay devices to forward messages, and to support edge servers. Moreover, through \textit{federated learning} mechanisms, UAVs can also help in processing the collected data and sharing the \textit{learned} model to the fog/cloud servers where all the received models are aggregated and compared for decision making. Adopting such an approach adds extra constraints on computation, which requires optimization of the UAV resources (CPU, battery, node election and formation) based on efficient task allocation and scheduling (sensing, communication and computing), in addition to many other mechanisms that help in reducing energy consumption and increase the operation lifetime.
    
    \item \textbf{Data type diversity}: drones are exchanging various types of messages that should be treated differently. Some of exchanged data is used to guide the drones to fulfill their tasks (i.e. message exchange between UAVs or instructions received from the pilot), IoT messages generated by the UAV and other messages generated from other devices and relayed by the flying drone.
    \item\textbf{System Infrastructure}: Deployment of the 5G infrastructure is under process in many countries and may take years before having a full 5G coverage.
    \item \textbf{Regulations and standards}: Regulation is a big challenge for both blockchain and UAVs. New rules should be defined to regulate the involvement of drones in cities  for safety and privacy reasons and to decide entities and service providers that may have access to the data in the blockchain. Part of these rules should regulate the relationship between the various service providers.
    \item \textbf{Novel UAV services}:the proposed system allows to achieve high automation and intelligence level that is the key to designing novel and advanced UAV services.
    \item \textbf{Innovative consensus algorithms}: advanced consensus algorithms should be designed, considering the various types of services provided by the aircraft. 
    \item \textbf{Off-chain blockchain storage}:As mentioned above, UAVs exchange various types of data. Some of the data may be too large to be stored in the blockchain efficiently or requires frequent modification or deletion. An off-chain blockchain storage should be provided to solve this issue and enhance the system performance.  
    \item \textbf{Dense number of nodes:} various challenges such as location identification, channel selection, interference-management and the availability of the line of sight.
    \item \textbf{Location and Divisibility}: One of the primary challenges for the proposed model is the mutual agreement on the location and divisibility of services between the service providers.
\end{itemize}

\section{Conclusion}
This article presented research guidelines for a 5G-UAV network that provides fast, reliable and secure service delivery to end-users of smart cities, namely, Drones as a Service (DaaS). With the aid of blockchain, UAVs can act as mobile access points, routing entities or resource providers in a decentralized manner. With the exponential growth in the number of connected IoT devices, data and service acquisition may be irregular depending on the location and time. Thus, it may be more beneficial both economically and in terms of service quality to run such a blockchain-assisted UAV network instead of a permanent fixed and over-provisioned cellular network. The solution supports service delivery either with the aid of a 5G network or in a totally decentralized manner using only UAVs. Both public and private blockchains are deployed with the UAVs. Moreover, with the support of fog and cloud computing resources, data and services are efficiently delivered. A case study was performed using a network simulator to compare the data delivery success rate and number of messages exchanged using the proposed solution and other state-of-the-art UAV-supported data delivery techniques.
\bibliographystyle{IEEEtran}
\bibliography{Ref}

\begin{IEEEbiography}[{\includegraphics[width=1in,height=1.25in,clip,keepaspectratio]{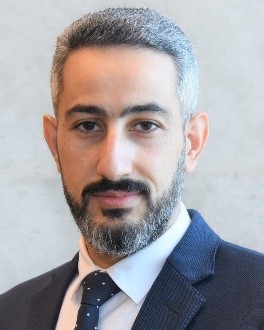}}]{Moayad Aloqaily (S’12, M’17)} received the Ph.D. degree in electrical and computer engineering from the University of Ottawa, Ottawa, ON, in 2016. He was an instructor in the Systems and Computer Engineering Department at Carleton University, Ottawa, Canada. He is working with Gnowit Inc.as a Senior Researcher and Data Scientist since 2016. He is also the managing director of xAnalytics Inc., Ottawa, ON, Canada. Currently, he is with the Faculty of Engineering, Al Ain University, United Arab Emirates. His current research interests include AI and ML, Connected and Autonomous Vehicles,Blockchain Solutions, and Sustainable Energy and Data Management. He is an IEEE member and actively working on different IEEE events. He has chaired and co-chaired many IEEE conferences and workshops such as BCCA2020,PEDISWESA-ISCC2020, ITCVT-NOMS2020, E2NIoT-IWCMC2020, ICCN-INFOCOM19,AICSSA’19, BAT-FMEC’19’20. He is a guest editor in many journals including International Journal of Machine Learning and Cybernetics, Elsevier IPM Journal, Springer JONS, Springer Cluster Computer, Internet Technology Letters,Transaction on Telecommunications Technologies, Security and Privacy, and IEEE Access. He is an Associate Editor with IEEE Access, Cluster Computing, Security and Privacy. He is a Professional Engineer Ontario (P.Eng.).
\end{IEEEbiography}

\begin{IEEEbiography}[{\includegraphics[width=1.25in,height=1.25in,clip,keepaspectratio]{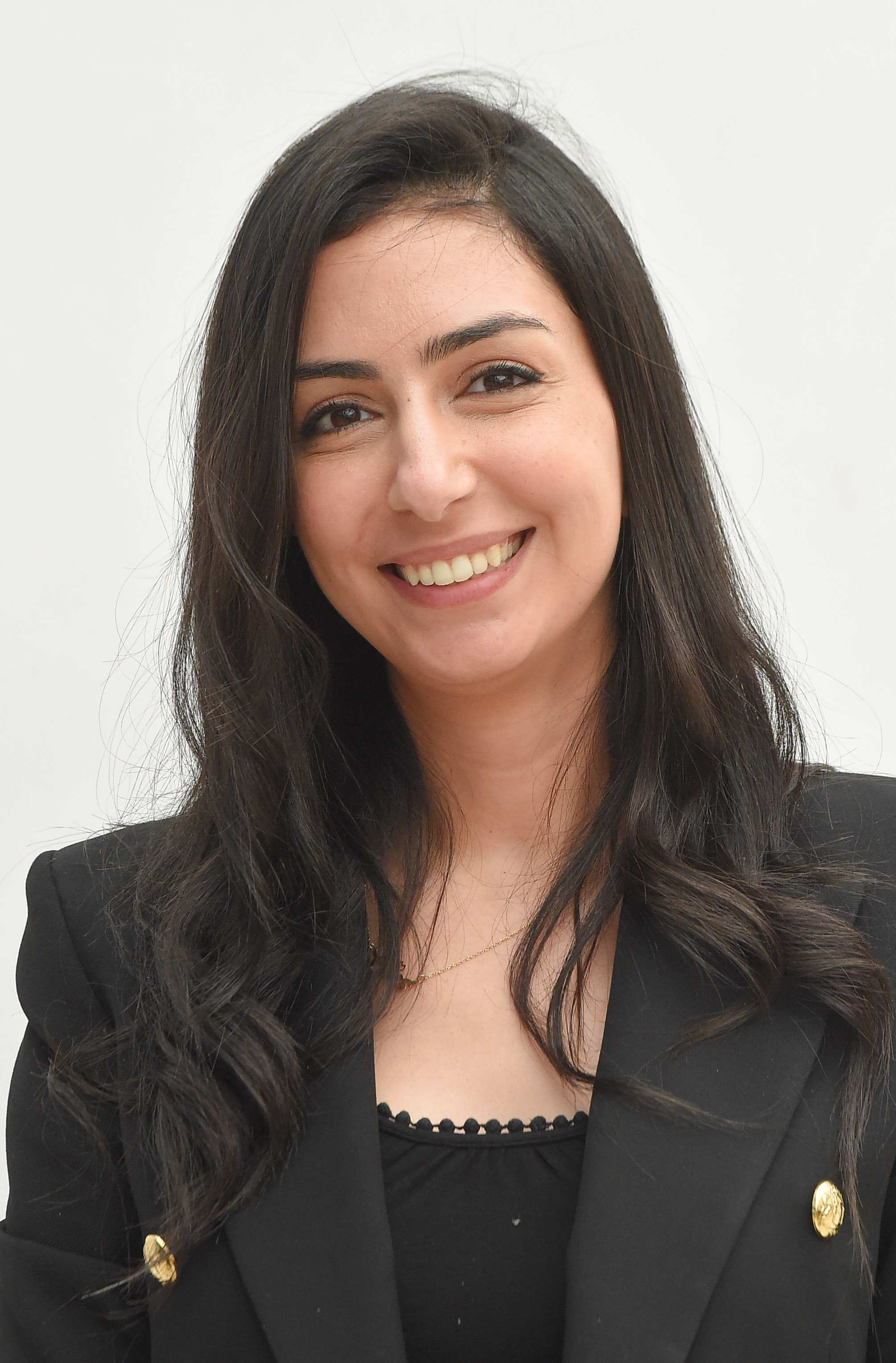}}]{Ouns Bouachir (M’18)} is an assistant professor of computer engineering in the college of Technological Innovation at Zayed University, UAE. She has a PhD degree in computer engineering from the University Paul Sabatier (Toulouse 3), Toulouse, France.  She also holds an engineering degree in Telecommunications from the Higher School of Communications (Sup’Com), Tunis, Tunisia. Prior to joining Zayed University, she worked as an assistant professor and post-doc researcher at Canadian University Dubai. Her current research activities are directed toward Connected and Autonomous Vehicles, Blockchain Solutions, Internet of Things (IoT), Energy Optimization, AI and ML. She is an IEEE member and part of the technical program committee for numerous journals, conferences and workshops like BCCA2020, ITCVT-NOMS2020, PEDISWESA-ISCC2020, and AICSSA2019.
\end{IEEEbiography}

\begin{IEEEbiography}[{\includegraphics[width=1in,height=1.5in,clip,keepaspectratio]{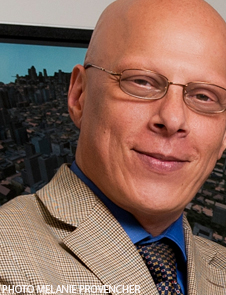}}]{Azzedine Boukerche (F’15)} is a Distinguished University Professor and holds a Canada Research Chair Tier-1 position at the University of Ottawa, Canada. He has received the C. Gotlieb Computer Medal Award, the Ontario Distinguished Researcher Award, the Premier of Ontario Research Excellence Award, the G. S. Glinski Award for Excellence in Research, the IEEE Computer Society Golden Core Award, the IEEE CS-Meritorious Award, the IEEE TCPP Technical Achievement and Leadership Award, the IEEE ComSoc ASHN Leadership and Contribution Award, and the University of Ottawa Award for Excellence in Research. His current research interests include wireless ad hoc and sensor networks, mobile computing, and performance evaluation of large-scale distributed and mobile systems. He has published extensively in these areas. He is a Fellow of the Engineering Institute of Canada, a Fellow of the Canadian Academy of Engineering, and a Fellow of the American Association for the Advancement of Science.
\end{IEEEbiography}

\begin{IEEEbiography}[{\includegraphics[width=1in,height=1.25in,clip,keepaspectratio]{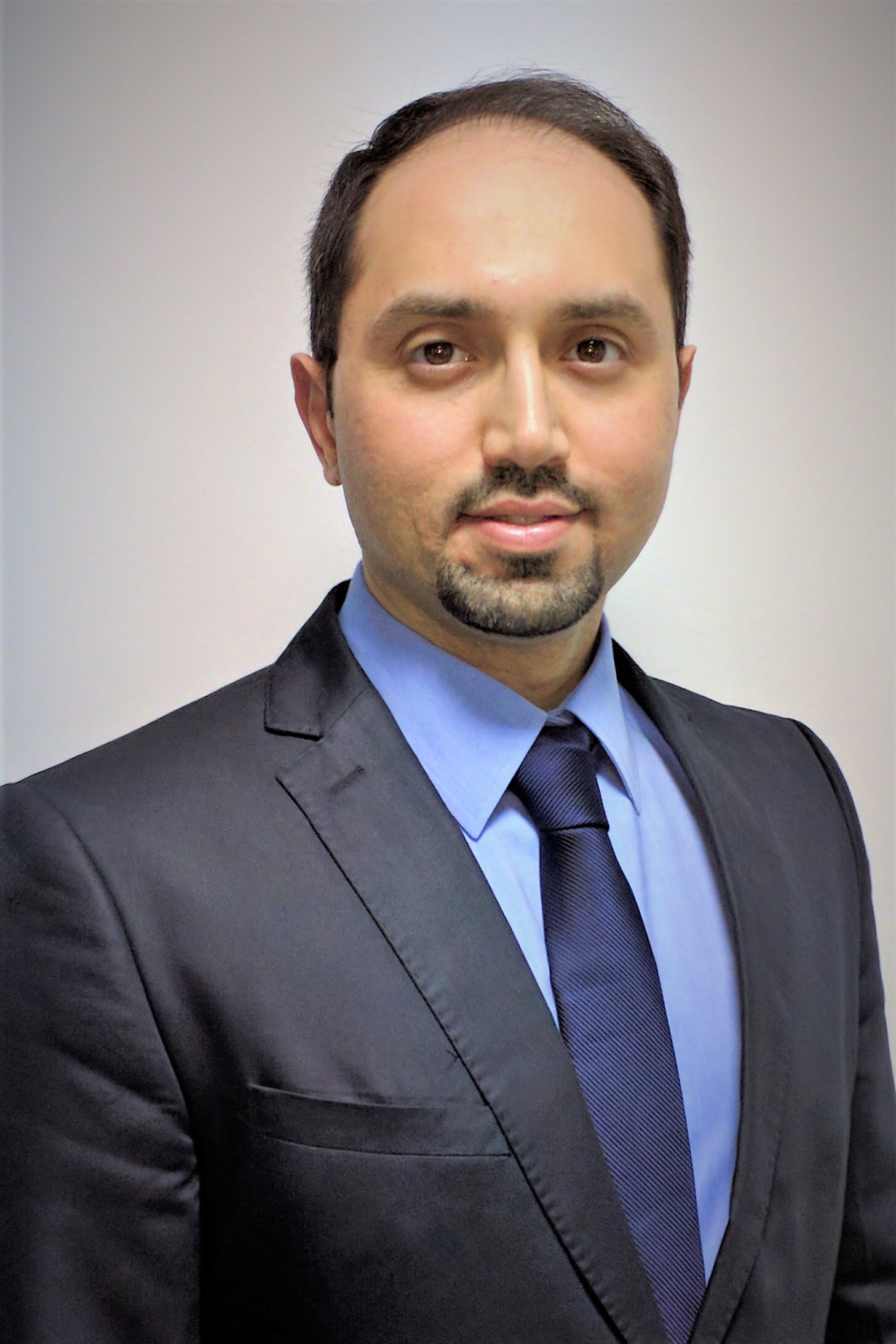}}]{Ismaeel Al Ridhawi (M’09, SM'19)} received his BASc, MASc, and Ph.D degrees in Electrical and Computer Engineering from the University of Ottawa, Canada, in 2007, 2009, and 2014 respectively. He is an Assistant Professor of Computer Engineering at Kuwait College of Science and technology and a researcher in the field of wireless communications. He has also worked at the American University of the Middle East in Kuwait as an Assistant Professor from 2014 to 2019. He is a registered professional engineer in Ontario (P.Eng). He's a Senior IEEE member with many peer-reviewed publications in highly ranked magazines, journals and conference proceedings. He is an associate and guest editor in many journals and has organized a number of IEEE conferences over the years. He has also served as session chair for a number of symposiums and was part of the technical program committee for numerous journals and conferences. His current research interests include service delivery and provisioning in fog and cloud computing, quality of service monitoring for wireless networks, MEC network management, and service-specific overlay networks.
\end{IEEEbiography}
\balance
\end{document}